# Experimental study of negative photoconductivity in $n$-PbTe(Ga) epitaxial films

Boris A. Akimov and Vladislav A. Bogoyavlenskiy
*Low Temperature Physics Department, Moscow State University, Moscow 119899, Russia*

Ludmila I. Ryabova
*Chemistry Department, Moscow State University, Moscow 119899, Russia*

Vyacheslav N. Vasil'kov
*Research Development and Production Center ORION, Moscow 111123, Russia*

We report on low-temperature photoconductivity (PC) in $n$-PbTe(Ga) epitaxial films prepared by the hot-wall technique on $\langle 111 \rangle$-BaF$_2$ substrates. Variation of the substrate temperature allowed us to change the resistivity of the films from $10^8$ down to $10^{-2}$ $\Omega$ cm at 4.2 K. The resistivity reduction is associated with a slight excess of Ga concentration, disturbing the Fermi level pinning within the energy gap of $n$-PbTe(Ga). PC has been measured under continuous and pulse illumination in the temperature range 4.2–300 K. For films of low resistivity, the photoresponse is composed of negative and positive parts. Recombination processes for both effects are characterized by nonexponential kinetics depending on the illumination pulse duration and intensity. Analysis of the PC transient proves that the negative photoconductivity cannot be explained in terms of nonequilibrium charge carriers spatial separation of due to band modulation. Experimental results are interpreted assuming the mixed valence of Ga in lead telluride and the formation of centers with a negative correlation energy. Specifics of the PC process is determined by the energy levels attributed to donor Ga$^{III}$, acceptor Ga$^{I}$, and neutral Ga$^{II}$ states with respect to the crystal surrounding. The energy level corresponding to the metastable state Ga$^{II}$ is supposed to occur above the conduction band bottom, providing fast recombination rates for the negative PC. The superposition of negative and positive PC's is considered to be dependent on the ratio of the densities of states corresponding to the donor and acceptor impurity centers.

## I. INTRODUCTION

Experimental data on persistent negative photoconductivity (PC) phenomena in semiconductors were reported in a number of papers dealing with quantum-well structures based on III-V (Refs. 1–3) and II-VI compounds.[4] In IV-VI semiconductors, negative PC was observed in polycrystalline films of nominally undoped $n$-PbTe,[5] in epitaxial films of Pb$_{1-X}$Sn$_X$Te(In), and in epitaxial films of $n$-PbTe with Pb excess.[6]

PbTe is well-known as a narrow-gap semiconductor sensitive in the 3–5 $\mu$m spectral region [middle infrared (IR)].[7] Four valleys forming a Fermi surface occur at the $L$ points of the Brillouin zone. High permittivity combined with low effective masses results in a high carrier mobility.[8] Photodetectors and thermal vision arrays with a sensitive element based on PbTe films were developed.[9] The main obstacle to optimizing the device parameters arises from the high carrier concentration due to electrically active native defects. Ga is the only impurity known to reduce carrier concentration in PbTe to nearly intrinsic values. Ga, as well as In and Tl, behaves as an impurity with a mixed valence in the lead telluride.[10] In contrast to species forming shallow levels and usually affecting only carrier concentration, mixed valence impurities in semiconductors can qualitatively modify properties of an initial material. $DX$ centers in III-V and II-VI semiconductors may be considered a classic example of mixed valence impurity behavior, providing the formation of impurity states with negative correlation energy $U$.[11]

The present state of the theory offers no possibility to predict correctly whether an impurity would reveal a mixed valence behavior in the lead telluride or its solid solutions. It was established experimentally that in addition to the group-III elements In, Tl, and Ga, the transition metal Cr (Refs. 12 and 13) and the rare-earth elements Yb (Refs 14 and 15) and Gd (Ref. 16) form impurity centers providing the Fermi level (FL) pinning. In contrast to the classic $DX$ centers, the FL in doped PbTe may be pinned in allowed bands as well as within the gap. $DX$ center formation in III-V compounds results in the appearance of a deep (ground) level and a shallow (excited) one. Study of a PC transient in lead telluride doped with Ga proved that the impurity levels associated with both the ground and metastable states of impurity centers are separated from the extended ones by a barrier,[17] so both impurity levels are deep.

Usually the mixed valence of an impurity together with the FL pinning result in the appearance of a prolonged saturation region that depends on the carrier concentrations $n$, and $p$ of the impurity content $N_i$. Ga-doped PbTe is an exception. Successive increases of $N_{Ga}$ in PbTe result in a p-n conversion and a narrow saturation region, followed by a further rapid increase of electron concentration.[18] $n$-PbTe(Ga) samples corresponding to the saturation region are semi-insulating at low temperatures due to the FL pinning within the gap at 70 meV below the conduction-band bottom. The persistent PC is observed in semi-insulating samples at temperatures $T < T_C$. In bulk material $T_C$ is about 80 K, while in films it reaches 110 K. For lead telluride doped with Ga, only positive PC was observed earlier. Photoelectric properties, transient processes,[17,19] PC spectra,[20] and magnetic susceptibility behavior[21] were studied mainly for semi-insulating samples of $n$-PbTe(Ga).

However, the microscopic theory of impurity states in PbTe(Ga) is still under discussion. Additional information necessary to develop the theory may be obtained through an investigation of $p$-PbTe(Ga) samples with doping levels insufficient to pin the FL, or of $n$-PbTe(Ga) samples where the impurity content slightly exceeds $N_{Ga}$, corresponding to the saturation region in the curve $n,p(N_{Ga})$. The first group of samples was found to be nonhomogeneous.[22] In this work, we report on a study of PC and transient processes in the second group of $n$-PbTe(Ga) samples with a relatively low resistivity compared to semi-insulating samples. The paper is organized as follows. In Sec. II we describe the sample preparation technique, the microstructure of the films grown, and details of low-temperature measurements. The subjects of Sec. III are experimental results of PC measured in darkness and under IR sources, and low-temperature transient processes. In Sec. IV we discuss the results obtained and propose a model, which allows us to explain the negative PC effect observed. Finally, Sec. V gives a summary of the paper.

## II. SAMPLES AND EXPERIMENTAL TECHNIQUE

The films were prepared by the hot wall technique on $BaF_2\langle 111\rangle$-oriented substrates. The initial growth melt consisted of 90% PbTe and 10% GaTe. The temperature of the evaporation source in the growth cell was fixed at 740 °C, and the substrate temperature $T_{sub}$ was varied from 160 °C up to 280 °C. The conductivity type and value depend on $T_{sub}$. At a value $T_{sub}$ lower than 200 °C the films were $p$ type. Semi-insulating $n$-PbTe(Ga) films were obtained in the interval $T_{sub}=210-240$ °C. Properties of these films are similar in their main features to semi-insulating bulk $n$-PbTe(Ga) single crystals, with the FL pinned within the gap. Photoelectric properties of these films were described in Ref. 17. Successive increases of $T_{sub}$ up to $\sim 280$ °C result in a conductivity enhancement, a significant reduction of the impurity activation energy, and a qualitative change of the PC character.

The thickness of the films was varied from 1 up to 3 $\mu$m depending on the synthesis duration. The PbTe(Ga) layers grown had good adhesion to the substrate, and a mirrorlike surface. The structure of as-grown films was controlled with use of electron microscopy, acoustic microscopy, and x-ray diffraction. Films were oriented along the $\langle 111\rangle$ direction. For all investigated films, x-ray reflections are characterized by high intensity, and the halfwidth of each reflection is given by the instrumental broadening. Surface morphology was studied with a TESLA BS-301 electron microscope and an ELSAM acoustic microscope. The distribution of the Ga impurity through the film volume was nonhomogeneous. The observed impurity segregations had a slight dispersion of their dimensions with a mean diameter of 0.5–1 $\mu$m, and the segregation density was about $10^6$ cm$^{-2}$. It should be mentioned that the dimensions of segregations in the films are an order of magnitude less than in the bulk samples of semi-insulating $n$-PbTe(Ga), but their density is an order of magnitude higher.[23]

An $In+4\% Ag+1\% Au$ alloy was used for the preparation of contacts. Low-temperature measurements of the conductivity were performed with use of a special cell screening a sample from background illumination. A thermal source and a GaAs light-emitting diode (LED) with $\lambda=1 \mu$m were used for IR illumination of the films. The maximal illumination flux density was $10^{-5}$ W/cm$^2$ for the thermal source and $10^{-4}$ W/cm$^2$ for the LED.

## III. EXPERIMENTAL RESULTS

### A. Temperature dependence of resistivity

Typical temperature dependences of the film resistivity $\rho$ in darkness ($\rho_{dark}$) and under continuous illumination by the thermal source are shown in Fig. 1. $\rho_{dark}(T)$ curves are characterized by a rather slight growth of resistivity at cooling from the room temperature down to $T\sim 50$ K. Further cooling results in saturation [curves 1 and 2 in Fig. 1(a)] or even in a decrease of the $\rho_{dark}$ value for the lowest resistivity sample [curve 3 in Fig. 1(a)]. The observed behavior of $\rho_{dark}(T)$ differs significantly from that of semi-insulating films, in which the ratio $\rho_{dark}^{4.2\ K}/\rho_{dark}^{300\ K}$ exceeds $10^6$, and the impurity activation energy is $E_A\sim 70$ meV.[17]

The photoresponse of the low resistivity $n$-PbTe(Ga) films is composed of negative (NPC) and positive (PPC) parts. Contributions of these parts to the PC process depend on the temperature and resistivity of a film in the darkness. The films are sensitive to IR illumination at temperatures $T_C<100-110$ K. The coexistence of the NPC and PPC processes does not allow us to determine amplitudes for each of the effects, but it is possible to make qualitative estimations. The temperature $T_C$ corresponding to the appearance of the PPC signal in the investigated films resembles the value obtained for semi-insulating $n$-PbTe(Ga) samples, where the amplitude of the PPC response increases at cooling and saturates at temperatures lower than 50 K. In the case of low-resistivity films, one more characteristic temperature $T_N$ corresponding to the beginning of the positive photoresponse reduction appears [Fig. 1(a)]. At $T=T_N$, contributions of the NPC and PPC to the photoresponse value become comparable. The temperature of intersection of the $\rho(T)$ curves obtained in the darkness and under illumination corresponds to equal amplitudes of the PPC and NPC processes. The subsequent reduction of temperature results in a monotonic decrease of the resistivity taken under illumination. Thus the amplitude of the NPC effect monotonically rises with cooling. The character of $\rho(T)$ curves obtained at different illumination fluxes proves this conclusion [Fig. 1(b)]. The amplitude of the NPC signal rises with the illumination intensity increase. Curve 4 in Fig. 1(b) was measured in the following way. First, the sample was illuminated by the thermal source at $T=4.2$ K. Then the source was switched off, and the $\rho(T)$ curve was taken at a temperature rise. This curve shows that the NPC is persistent, and its amplitude exceeds the amplitude of the persistent PPC through the whole temperature interval $4.2<T<T_C$. It is also important to note that the lower the resistivity of the film, the more clearly the NPC effect can be seen.

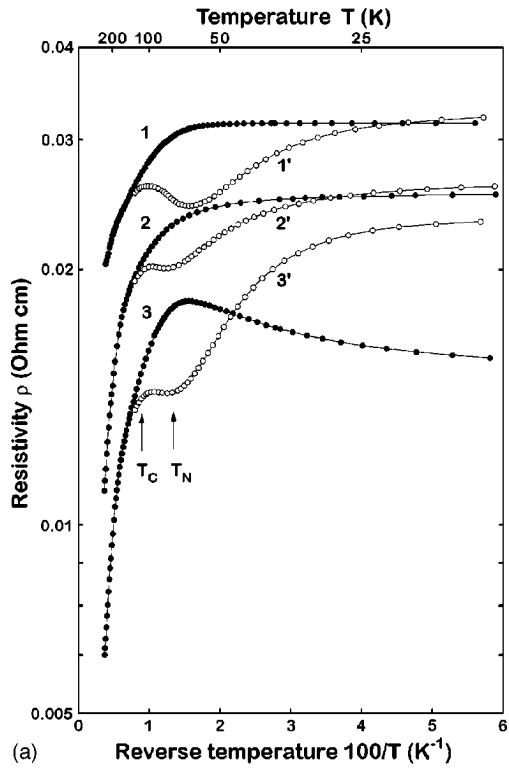

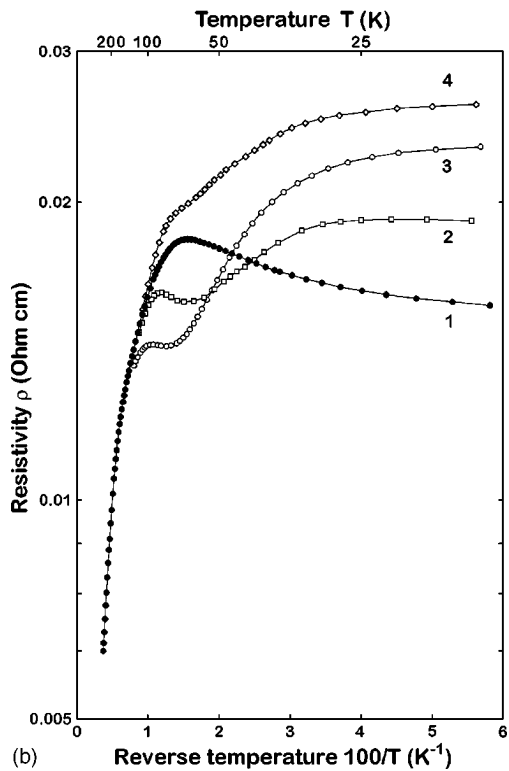

FIG. 1. Temperature dependence of the resistivity $\rho$ for $n$-PbTe(Ga) films with slightly different conductivity values; (a) curves 1–3 were measured in darkness, curves $1'$–$3'$ under illumination by a thermal source; (b) curves 1–4 show the typical behavior of the $\rho(T)$ dependence for one of the films measured: curve 1 at cooling in darkness, curves 2 and 3 at heating from the liquid helium temperature under continuous illumination by the thermal source with different intensities (illumination intensity is higher for the curve 3), and curve 4 at heating in darkness after the film was illuminated by the thermal source at $T=4.2$ K.

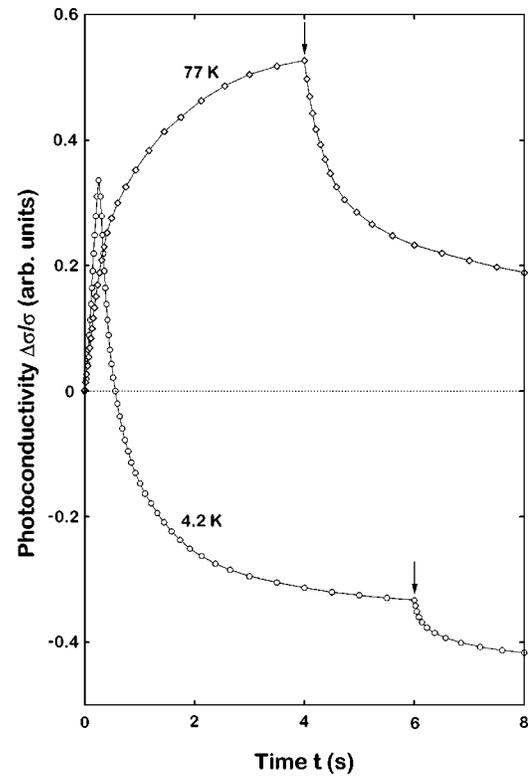

FIG. 2. Photoconductivity transient measured under illumination by the thermal source in $n$-PbTe(Ga) films at 77 K and 4.2 K. Arrows mark the moments the illumination is switched off.

### B. Photoconductivity transient

The results of a transient process investigation under continuous illumination by the thermal source at different temperatures are presented in Fig. 2. The curve taken at liquid nitrogen temperature has no qualitative distinction from the PPC transient curves obtained for semi-insulating $n$-PbTe(Ga) films.[17] This result shows that the superposition of the NPC and PPC effects strongly depends on the order of the experimental procedure. If the sample is illuminated at liquid-helium temperature, the NPC persists up to temperatures of about $T_C$ [curve 4 in Fig. 1(b)], but illumination of the sample in the temperature region of PPC domination does not reveal the NPC processes. The photoresponse transient at liquid-helium temperature reveals a sharp peak of the PPC occured after the light is switched on. The transient drastically changes to NPC in 0.2–0.4 s (the decrease of the PC under IR illumination). The appearance of an additional rapid decrease of the conductivity after the illumination is switched off demonstrates the contribution of the PPC to the PC process at a relatively quick characteristic time.

Pictures illustrating transient processes induced by LED illumination are presented in Fig. 3. Photoresponse under successive illumination pulses (the pulse duration is $5\times10^{-6}$ s) at $T=4.2$ K is shown in the inset. In contrast to the transient under continuous illumination, LED pulses induce only NPC, and no persistent effects appear. The shape of the relative photoresponse $\Delta\sigma/\sigma$ corresponding to one pulse of illumination is shown on a linear plot [Fig. 3(a)] and on a logarithmic plot [Fig. 3(b)] at different temperatures. The amplitude of the NPC response reduces smoothly with a temperature increase. The relaxation process is characterized by nonexponential kinetics. The characteristic relaxation time $\tau$ defined as $\tau = -\Delta\sigma/(\partial\sigma/\partial t)$ is about $10^{-5}$ s, just after the illumination pulse, and its change with temperature is negligible. The value of $\tau$ rises during the relaxation, and the rate of a $\tau$ increase drops with the temperature rise.

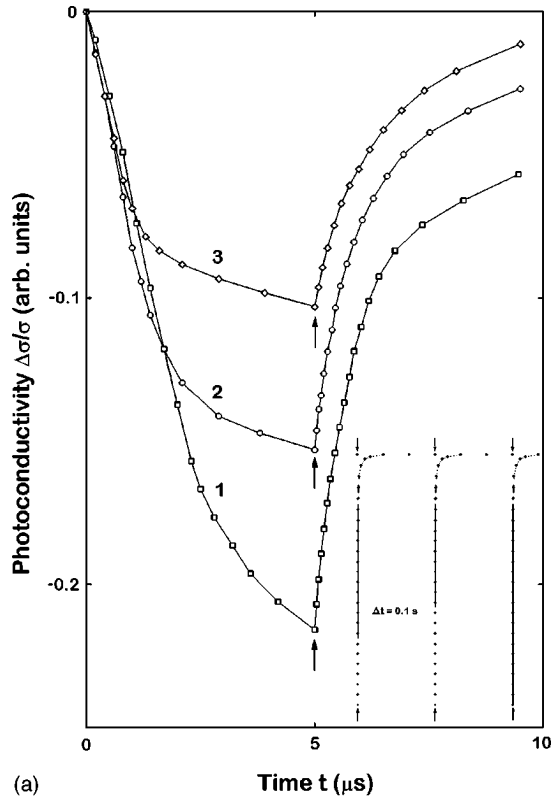

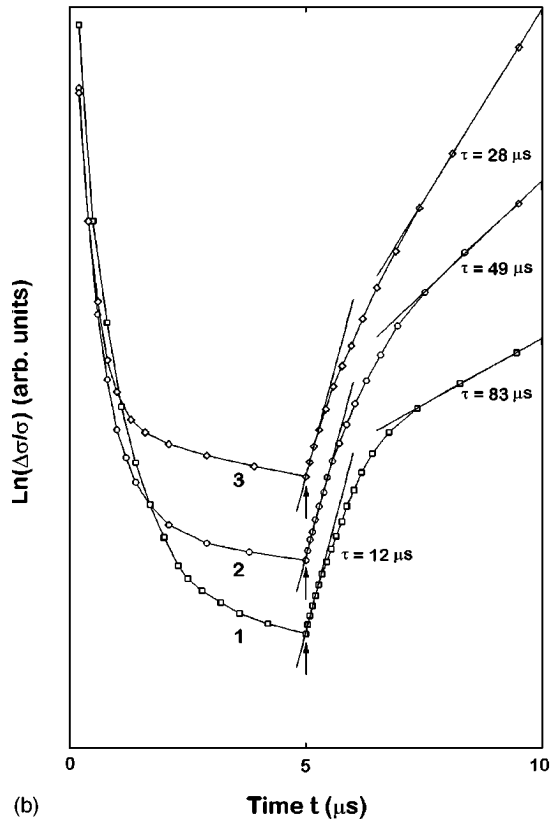

FIG. 3. Photoconductivity transient measured under LED pulses (pulse duration is 5 $\mu$s) in $n$-PbTe(Ga) films at 4.2 K (curve 1), 13 K (curve 2), and 30 K (curve 3) in a $\Delta\sigma/\sigma$ vs $t$ plot (a) and a $\ln(\Delta\sigma/\sigma)$ vs $t$ plot (b). Arrows mark the moments the illumination is switched off. The insert shows curve 1 on a prolonged time scale.

## IV. DISCUSSION

### A. Negative photoconductivity in lead telluride and related materials

The most important feature of the NPC in the investigated $n$-PbTe(Ga) films is the appearance of a fast region in transients under LED pulses. In contrast to the PC transient in quantum-well III-V and II-VI structures, this fast process, with a characteristic time of the order of dozens of microseconds, can hardly be explained by band modulation and spatial separation of nonequilibrium carriers. The NPC transient in polycrystalline PbTe films also differs significantly from that observed in $n$-PbTe(Ga).[5] The NPC observed in polycrystalline films at $T>77$ K reaches a maximum at 85 K, so the amplitude of the negative photoresponse is nonmonotonic with temperature. In addition, the NPC transient is nonexponential, with characteristic times of about 10 s ($T=77$ K) at the beginning of the relaxation, followed by the long-term relaxation tail. The PC character of $n$-PbTe films grown with an excess of Pb (Ref. 6) correlates better with the NPC in polycrystalline films than with the NPC in $n$-PbTe(Ga) films. The NPC appears at 60 K. Its characteristic time $\tau$ is about 0.5 s at $T=60$ K. $\tau$ rapidly reduces with the temperature increase, obeying the relation $\tau = \tau_0 \exp(W/kT)$, where $W \sim 85$ meV for $T=100$–170 K. The NPC was observed up to room temperature, but its amplitude reached maximum at $T=165$ K. Recombination of the NPC on surface traps in $n$-PbTe films was also found to be of a long duration: the characteristic time is of the order of hours at $T=77$ K.[24] The fast NPC with characteristic times of $10^{-3}$–$10^{-4}$ s was observed earlier in $Pb_{1-x}Sn_xTe(In)$ films with $0.19<x<0.23$.[6] The effect was registered under illumination of a laser combined with background radiation of a sample. In contrast to our results, the NPC in $Pb_{1-x}Sn_xTe(In)$ films is not absolute: it is rather a quenching of the persistent PPC induced by background illumination.

### B. Energy spectrum of impurity states in PbTe(Ga): theory and experiment

#### 1. Mixed-valence behavior of Ga impurity

Existence of three valent states of Ga follows from the possibility to find these states in individual chemical compounds of Ga, e.g., in halides $GaX_n$ where $X=Cl, Br,$ and I. It is well known, e.g., that the compounds GaCl, $GaCl_2$, and $GaCl_3$ exist (the second one is metastable and decomposes into GaCl and $GaCl_3$). In these compounds, the number of halogen atoms surrounding Ga plays the role of a parameter that gives rise to the valence change. As a result, the energy gain due to the formation of an additional bond in the molecule compensates for the energy loss due to the change in the electronic configuration of the group-III atom. In a semiconductor, however, the number of atoms surrounding an impurity center does not change when the valence of impurity changes. To explain the valence change of an impurity atom in a semiconductor, the concept of crystal-lattice distortion nearby the impurity center is usually suggested. In particular, this approach was used to explain the behavior of DX centers in III-V and II-VI semiconductors.

Most theoretical models describing group-III impurity behavior in lead telluride and related materials, while varying in details assume a negative-U behavior of a center with two localized electrons. According to Ref. 25, the negative-U behavior of a Ga impurity center leads to the dissociation of $Ga^{II}$ states which are neutral with respect to crystal surroundings, into donor $Ga^{III}$ states and acceptor $Ga^{I}$ states. The reaction

$$2Ga^{II} \rightarrow Ga^{III} + Ga^{I} \qquad (1)$$

provides the FL pinning. The effective attraction of two electrons localized at the acceptor $Ga^{I}$ center leads to the reduction of its energy. Therefore, the energy corresponding to the ground $Ga^{I}$ state is lower than the energy of the metastable $Ga^{II}$ state with one localized electron.

Direct experimental observations prove the existence of two impurity levels in semi-insulating n-PbTe(Ga). Thermal activation of the impurity conductivity corresponds to an energy level lying at about 70 meV below the conduction-band bottom.[10] This level is associated with the ground state responsible for the FL pinning. One more energy level nearby the conduction-band bottom was observed in a PC spectra study.[19,20] A character of the PPC transient may also be understood by assuming two energy levels separated from the extended states by energy barriers of different heights. These results may be qualitatively explained in terms of the dissociation reaction [Eq. (1)], and by crystal lattice distortion nearby an impurity atom when its charge state is changed. Nevertheless, it was rather difficult to define the electronic configuration of Ga impurity centers considering the photoconductivity effect only.

### 2. Microscopic theory

A theoretical approach defining the electronic configurations of Ga impurity centers was developed in Refs. 26 and 27 to explain the sharp paramagnetic peak at 60 K observed in low-temperature magnetic susceptibility measurements of semi-insulating n-PbTe(Ga) crystals.[21] The model is based on an idea of the closeness of the total (multielectronic) energies of different electronic configurations ($s^2p^1, s^1p^2$, and $s^0p^3$) of an impurity atom substituting for Pb at the site of a crystal lattice of a semiconductor. Metal atoms in IV-VI semiconductors are in a divalent state ($s^2p^2$ configuration). Their s electrons form deep, completely filled bands, and the atomic p orbitals form the valence and conduction bands. It is therefore obvious that the group-III atom Ga substituting for a divalent metal Pb in the ground (univalent $s^2p^1$) atomic configuration is a singly charged acceptor; in the divalent configuration ($s^1p^2$) an impurity atom is neutral and paramagnetic (unpaired spin in the s shell), and in a trivalent configuration ($s^0p^3$) it becomes a singly charged donor. The valence changes at the moment when the thermodynamic potentials of the two configurations $s^2p^1$ and $s^0p^3$ are equal. According to the experimental data, the pinning occurs in PbTe(Ga) within the band gap. When the valence changes, electrons delocalize. It was assumed that, to a good degree of accuracy, the thermodynamic potential of the valence and conduction bands changes only as a result of a change in the occupancy near the L points. The conditions for pinning were determined by studying the thermodynamic potential of the system ''s level + L point.''

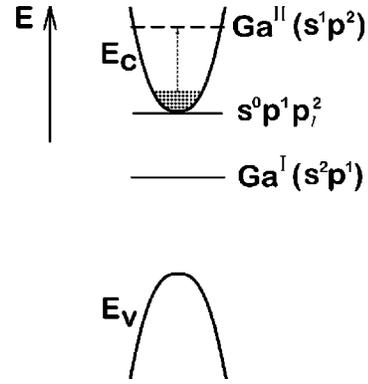

FIG. 4. Diagram illustrating the energy spectrum and impurity states in n-PbTe(Ga).

One more conclusion of the approach mentioned above is the following. If an impurity atom in a semiconductor is in a nondominant configuration, an attractive potential arises for electrons in the conduction band due to the interaction of the electrons with deep holes localized on the s level of an impurity. When this interaction is sufficiently strong, a local state $s^0p^1p_l^2$ splits from the bottom of the conduction band at the L point. The energy of the split-off level was estimated as 5–10 meV.[26] Thus this energy appears to be quite close to the energy of a level experimentally observed in PC spectra.

### 3. Photoconductivity phenomena

One of the basic suggestions of Refs. 26 and 27 is that electrons of the Ga p shell participate in the formation of actual bands. Therefore, photoionization of an impurity atom provides a redistribution of electrons between its s and p shells.[27,28] The ground state of the impurity $s^2p^1$ reveals a negative-U behavior; thus the metastable energy level corresponding to the neutral with respect to the crystal surrounding $s^1p^2$ configuration is higher than the FL, and may occur above the conduction-band bottom [see the scheme shown in Fig. 4]. Photoexcitation of the first electron from the univalent ground $s^2p^1$ state into the conduction band needs additional energy to transform the impurity center in the divalent state $s^1p^2$. This additional energy (equal to the energy distance between the conduction band bottom and the metastable level position) is similar to the recombination barrier in the models of lattice distortion nearby the impurity center when its charge state is changed. To explain the fast PPC, localization and delocalization of electrons in a short-range potential of an empty Ga s shell (a split-off level) are considered [Fig. 4].[28]

In low-resistivity films of n-PbTe(Ga), free electrons already exist in the conduction band. Simultaneous capture of two electrons is of a low probability. The NPC process may be associated with successive transitions of two free electrons into the s shell of a Ga impurity according to the process

$$s^0p^3 \rightarrow s^1p^2 \rightarrow s^2p^1. \qquad (2)$$

The capture of a free electron needs additional energy to transform the impurity center into the $Ga^{II}$ state (Fig. 4), and illumination may initiate this process. The persistent NPC may be understood by assuming localization of a second electron in the $s$ shell of an impurity center, resulting in a reduction of its energy. The fast NPC region may be related to metastable $s^1p^2$ state contributions. The metastable states' existence was also accepted in Ref. 21, where the Curie-Veiss behavior of a magnetic susceptibility $\chi$ at low temperatures was attributed to Ga centers in the $s^1p^2$ configuration.

However, the characteristic time of the fast NPC process is at least two orders of magnitude larger than the band-to-band recombination relaxation time, so the metastable states lying above the conduction-band bottom should be comparatively long living. This experimental result can hardly be explained by neglecting crystal distortion not specified in Ref. 26. Finally, it should be mentioned that long-term processes and even slow Fermi-surface reduction during relaxation of nonequilibrium carriers at levels lying in the bands continuum are rather usual phenomena for In doped PbTe and related materials, with the FL pinned within the allowed bands.[10]

### 4. Superposition of positive and negative photoconductivity

The superposition of PPC and NPC processes in $n$-PbTe(Ga) films may be related to the redistribution of a Ga impurity between donor and acceptor states. The quantity of donor atoms $Ga^{III}$ in low-resistivity films may be significantly larger than the quantity of $Ga^I$ atoms with two localized electrons. Thus the probability of electron capture into the $Ga^{III}$ state increases, while the probability of excitation of an electron from a $Ga^I$ state occupied by two electrons reduces. Therefore the NPC process becomes dominating. The coexistence of NPC and PPC processes cannot be excluded in semi-insulating samples either. As the PPC in this case dominates in the whole temperature region, this assumption cannot be proved by direct measurements. One of the arguments proving this suggestion is the stabilization of the conductivity value in illuminated samples at temperatures below 50 K, though the temperature dependence of the relaxation time (and, respectively, the conductivity) is exponential in the interval 50 K$<T<T_C$. This stabilization region may occur when the concentration of nonequilibrium carriers accumulated in the conduction band due to the PPC process, and the concentration of empty Ga centers, reach values corresponding to the balance between the NPC and PPC effects. That balance prevents a subsequent increase of conductivity. Thus for this case the NPC may be regarded as a partial quench of the PPC signal.

## V. CONCLUSIONS

The resistivity of PbTe(Ga) films grown by means of the hot wall technique depends strongly on the substrate temperature $T_{sub}$. An increase of $T_{sub}$ results in successive $p$-$n$ conversion, an increase of resistivity corresponding to the FL pinning within the energy gap, and a reduction of the resistivity. Semi-insulating $n$-PbTe(Ga) samples exhibit positive PC in the temperature range 4.2–110 K. Low-resistivity $n$-PbTe(Ga) films are characterized by the appearance of negative PC coexisting with positive PC. The negative PC dominates at temperatures lower than 20–40 K. Specifics of positive and negative PC transients can be understood in terms of generation-recombination processes in the system of impurity levels related to mixed valence of Ga in PbTe. Observation of the negative PC effect gives one more argument proving the existence of three impurity energy levels in the energy spectrum of $n$-PbTe(Ga) with the metastable energy level above the conduction-band bottom.

## ACKNOWLEDGMENTS

The authors would like to thank Professor D. R. Khokhlov for helpful discussions and Professor S. P. Zimin and Dr. M. N. Preobrazhensky for help in the PbTe(Ga) morphology study. This work was financially supported by RFBR under Project Nos. 98-02-17317 and 99-02-17531, and INTAS-RFBR under Project No. 95-1136.